# Observation of exceptional line semimetal in three-dimensional non-Hermitian phononic crystals

Yejian Hu[1†], Jien Wu[2†], Peidong Ye[1], Weiyin Deng[1,2*], Jiuyang Lu[1,2], Xueqin Huang[2], Ziyu Wang[1*], Manzhu Ke[1*], and Zhengyou Liu[1,3*]

[1]Key Laboratory of Artificial Micro- and Nanostructures of Ministry of Education and School of Physics and Technology, Wuhan University, Wuhan 430072, China
[2]School of Physics and Optoelectronics, South China University of Technology, Guangzhou, Guangdong 510640, China
[3]Institute for Advanced Studies, Wuhan University, Wuhan 430072, China

†These authors contributed equally to this work.
*Corresponding author.
Emails: dengwy@whu.edu.cn; zywang@whu.edu.cn; mzke@whu.edu.cn; zyliu@whu.edu.cn

**Non-Hermitian topological phases, which exhibit unique features such as skin effect and exceptional points originated from nontrivial band topologies in complex plane, have attracted enormous attention in condensed-matter physics and metamaterials. Here we report the realization of an exceptional line semimetal in a three-dimensional non-Hermitian phononic crystal. A pair of exceptional rings with opposite topologies are connected by the drumhead bulk states in the first Brillouin zone. The exceptional rings not only possess wave-function topology and thus result in the drumhead surface states, but also host spectral topology and thereby give rise to the hybrid-order geometry-dependent skin effect in three dimensions. Our experimental results evidence the complete non-Hermitian bulk-boundary correspondence of the three-dimensional exceptional line semimetal, and may pave the way for designing non-Hermitian acoustic devices.**

There has been growing interest in exploring topological phases of non-Hermitian systems, which host nontrivial band topologies in complex plane [1,2]. Among them, the topological semimetals with non-Hermitian band degeneracies, i.e., exceptional points whose eigenvectors also coalesce, are unique in the non-Hermitian systems [3-5]. This is because a point gap is open but a line gap is closed round the exceptional point, leading to the non-Hermitian Hamiltonian cannot be continuously deformed into the Hermitian or anti-Hermitian one [5,6]. The exceptional points are stable in the two-dimensional (2D) non-Hermitian systems, and a line of exceptional points, namely, exceptional line, is robust against weak perturbations without any symmetries in the 3D non-Hermitian systems [7,8]. A simple configuration of exceptional line in the 3D exceptional line semimetal is exceptional ring (ER), which usually can be induced in two ways [9-14]. One is introducing a non-Hermitian perturbation into the Weyl semimetal, deforming a Weyl point into a Weyl ER, which features with two topological charges described by a Chern number and a winding number [10-13]. Another is adding a non-Hermitian term of gain and loss into the nodal line semimetal, which splits a nodal ring into a pair of ERs only with the quantized winding number [14].

Based on the generalized Brillouin zone (BZ), the non-Hermitian bulk-boundary correspondence has twofold meanings [15]. One is that the conventional boundary states correspond to the wave-function topology of the non-Bloch Hamiltonian [16,17]. The other is that the non-Hermitian skin modes without a Hermitian counterpart, which permit the bulk states collapse to the open boundaries, correspond to the energy spectra topology of the Bloch Hamiltonian [18-21]. The 1D skin effect has been observed in different non-Hermitian wave systems, such as optical system [22] and phononic crystal (PC) [23]. The skin effects in two dimensions, including the corner skin effect [24-28] and geometry-dependent skin effect (GDSE) [29-32], have also been theoretically proposed and experimentally achieved. So far, the skin effect of 3D bulk states is rarely explored [32-34]. Due to the nontrivial wave-function topology with nonzero Chern number inherited from the Weyl point, the Weyl exceptional line semimetal exhibits the Fermi-arc surface states [10-13]. And naturally, the exceptional line semimetal induced from the nodal line semimetal has the drumhead surface states (DSSs) inherited from the nodal line [14]. However, the skin effect from the spectral topology of ERs in the exceptional line semimetal is still elusive. As a result, the exceptional line semimetal with the comprehensive non-Hermitian bulk-boundary correspondence is yet to be fully explored in the 3D systems.

In this work, we realize an exceptional line semimetal in a 3D non-Hermitian PC, based on a Hermitian PC of nodal ring semimetal. The PC at macroscopic scale for acoustic waves provides an ideal platform to explore frontiers of physics [35,36]. The non-Hermiticity is induced by the designed loss added to the coupling waveguide, and splits a nodal ring into a pair of ERs connected by the 3D drumhead bulk states in the first BZ. The ERs possess two types of topologies simultaneously. One is the wave-function topology characterized by the wave-function winding number, giving rise to the 2D DSSs. The other is the spectral topology revealed by the spectral winding number or spectral area, which can guarantee the hybrid-order GDSE in three dimensions. These topological properties are first illustrated in a tight-binding model, and then experimentally confirmed in the PCs by measuring the band dispersions and pressure field distributions. All the results of theories, simulations and experiments are consistent well with each other.

**Exceptional line semimetal in the tight-binding model**

We construct a 3D tight-binding model on a face-centered-cubic lattice, which is formed by a series of dimeric chains along the $z$ direction staggered on the $x$-$y$ plane, as shown in Fig. 1a. A uniform intralayer coupling $t_0$ is applied in the $x$ and $y$ directions, and the interlayer couplings $t_1$ and $t_2 - i\gamma$ exist in the $z$ direction, where $\gamma$ is the strength of designed loss to achieve the non-Hermiticity. The non-Hermitian Hamiltonian in momentum space has the form

$$H = (d_x - i\gamma \cos k_z)\sigma_x + (d_y - i\gamma \sin k_z)\sigma_y, \quad (1)$$

where $d_x = 2t_0(\cos k_x + \cos k_y) + (t_1 + t_2)\cos k_z$, $d_y = (t_2 - t_1)\sin k_z$, and $k_i$ and $\sigma_i$ ($i = x, y, z$) are the dimensionless wave vectors and Pauli matrixes denoting sublattice degree of freedom, respectively. The distance between the nearest-neighbor sites is set to unity. Figure 1b shows the first BZ denoted by the green truncated octahedron. For the Hermitian case $\gamma = 0$, there is a degenerate nodal ring (black dashed curve) on the $k_z = 0$ plane, which is protected by the mirror symmetry on this plane [37], forming the nodal line semimetal. The non-Hermitian interaction with $\gamma \neq 0$ makes the nodal ring split into a pair of ERs with $\pm 1$ spectral winding numbers (red and blue solid curves) connected by drumhead bulk states (yellow area), giving rise to the exceptional line semimetal. This is a 3D generalization of the 2D exceptional point semimetal, where a pair of exceptional points with opposite charges are connected by bulk Fermi arc [38]. Figures 1c and 1d show the real part of bulk dispersion and isofrequency curve of zero energy at $k_y = 0$ plane of the first BZ enclosed by orange dashed lines in Fig. 1b, respectively. A cross section of ERs (red and blue dots) and drumhead bulk states (yellow lines) can be seen. The detailed derivation and figure of ERs and drumhead bulk states are discussed in Methods and Supplementary S-I A with Fig. S1, respectively.

The ERs possess wave-function topology and spectral topology simultaneously. The wave-function topology of ERs is governed by the wave-function winding number [15] calculated by the non-Bloch wave function in the generalized BZ, as $w = \frac{i}{2\pi} \int_{C_\beta} q^{-1} dq$, where $q$ is formed by the right and left eigenvectors, and $C_\beta$ represents the generalized BZ for fixed $k_x$ and $k_y$. The calculated wave-function winding number on the $k_x$-$k_y$ plane is show in Fig. 1e. Red area, whose boundaries are consistent with the projections of ERs (gold line in Fig. 1e), shows wave-function winding number $w = 1$, and supports the existence of degenerate DSSs. The spectral topology of ERs is characterized by the spectral winding number [32] calculated by the energy spectra in the BZ, as $\nu(\mathbf{k}_{\mathrm{ER}}) = \frac{1}{2\pi i} \oint_{C_{\mathrm{ER}}} d\mathbf{k} \cdot \boldsymbol{\nabla}_{\mathbf{k}} \ln \det[H(\mathbf{k}) - E(\mathbf{k}_{\mathrm{ER}})]$, where $C_{\mathrm{ER}}$ represents a closed path enclosing the ER at $\mathbf{k}_{\mathrm{ER}}$ in momentum space anticlockwise and $E(\mathbf{k}_{\mathrm{ER}})$ is the corresponding eigenvalue. For example, energy spectra along the path shown in Fig. 1d, with $\theta$ varying from 0 to $2\pi$, exhibit a clockwise loop on the complex plane (Fig. 1f), revealing the spectral winding number $\nu = -1$. As such, the pair of ERs (red and blue solid curves in Fig. 1b) possess $\pm 1$ spectral winding numbers. These $\pm 1$ spectral winding numbers of ERs lead to the nonzero spectral winding number along some directions in the BZ, and eventually guarantee nonzero spectral area on the complex plane, resulting in the 3D GDSE [31,32]. Detailed calculations of topologies for ERs are discussed in Supplementary S-I B and C with Figs. S2 and S3.

The wave-function topology of ERs gives rise to the DSSs. We construct a ribbon sample under open boundaries in the $z$ direction and periodic boundaries in the $x$ and $y$ directions. Projected dispersion is calculated by the ribbon sample with 100 unit cells, in which the real and imaginary parts of drumhead surface dispersions are extracted and plotted in Figs. 2a and 2b, respectively. Flat dispersions at zero energy emerge from the nontrivial area with $w = 1$ (Fig. 1e), which are the DSSs localized at the open boundaries. Here the DSSs are similar to those in nodal line semimetals, but they originate from the wave-function topology of ERs rather than nodal rings. Although the DSSs exist at the open boundaries in the $z$ direction, they disappear at the open boundaries in the $x$ and $y$ directions because of the trivial wave-function winding number $w = 0$ in these directions.

Spectral topology of ERs can guarantee the GDSE in three dimensions. Finite-size samples under different open boundaries are considered to investigate the 3D GDSE, including cuboid-shaped geometry and pyramid-shaped geometry. A direct way to determine whether there is skin effect is to compare the spectral areas of sample under open and periodic boundaries, where the skin effect occurs if there is a difference between them [31,32]. Following this way, we plot the spectra (green dots) of the cuboid-shaped and pyramid-shaped geometries in Figs. 2c and 2e, respectively, where gray dots are spectra under periodic boundaries. One can see that the spectral area under cuboid-shaped geometry is the same as that under periodic boundaries, and hence does not support skin effect. On the contrary, the spectral area under pyramid-shaped geometry is different from that under periodic boundaries, giving rise to skin effect. The skin effect can be visualized by spatial distribution of eigenstates defined as $W(j) = \frac{1}{N}\sum_n |\psi_n(j)|^2$, where $\psi_n(j)$ is the $n$-th normalized right eigenstate at site $j$, and $N$ denotes the total number of eigenstates. Distributions of $W(j)$ under cuboid-shaped and pyramid-shaped geometries are shown in Figs. 2d and 2f, respectively. Skin effect only exists in the pyramid-shaped geometry, but not in the cuboid-shaped geometry, indicating this skin effect is GDSE. The GDSE in other cases is discussed in Supplementary S-I D-G with Figs. S4-S6. It is worth noted that the GDSE at one surface is determined by the nonzero spectral winding number along the normal direction of that surface. In this system, mirror symmetries guarantee zero spectral winding number along the $x$, $y$ and $z$ directions, leading to the disappearance of GDSE at surfaces perpendicular to these directions. We further reveal that the GDSE is a hybrid-order skin effect, including the surface and hinge skin modes, which can be predicted by the secondary deformation of the spectral area (Supplementary S-I H with Fig. S7).

**ERs and drumhead bulk states in the 3D non-Hermitian PC**

We realize the lattice model in the 3D non-Hermitian PC, where the acoustic cavities and coupling waveguides are functioned as the sites and couplings, respectively. Figure 3a shows the photograph of cuboid-shaped PC, which is formed by stacking multiple layers along the $z$ direction. The designed loss is achieved by the holes on the bigger interlayer waveguide sealed with the sound-absorbing sponges (black). Schematic of PC unit cell is shown in Fig. 3b, where acoustic cavities have width $d_0 = 15$ mm and height $h_0 = 30$ mm. Distance between nearest-neighbor cavities is $a = 30$ mm in the $x$ and $y$ directions and $h = 45$ mm in the $z$ direction. Intralayer waveguides in the $x$ and $y$ directions are the same and composed of two stacked cuboids: one has the width $d_1 = 9.5$ mm and height $h_1 = 1$ mm and another has the width $d_2 = 3.5$ mm and height $h_2 = 2.5$ mm. Bigger interlayer waveguide with width $d_4 = 6$ mm and small interlayer waveguide with width $d_3 = 4$ mm are applied in the $z$ direction. Green areas with width $d_4 = 6$ mm and height $h_3 = 4.5$ mm represent two rectangular holes on the bigger interlayer waveguide, which are used to achieve designed loss in experiment (Supplementary S-II with Fig. S8).

ERs and drumhead bulk states are experimentally confirmed in the cuboid-shaped PC sample, which contains 13 layers, and each layer has $13 \times 13$ cavities, as shown in Fig. 3a. Figure 3c shows measured (color map) and simulated (colored curves) isofrequency curves at 5600 Hz on the $k_z = \pm\pi/(6h)$, 0 planes, displaying ERs and partial drumhead bulk states, respectively. And the measured and simulated isofrequency curves at 5600 Hz on the $k_y = 0$ plane are plotted in Fig. 3d, revealing the drumhead bulk states terminated by ERs with opposite spectral winding numbers, which are calculated in Supplementary S-III with Fig. S9 and the same as those in the lattice model. Furthermore, we also measure (color map) and simulate (gray curves) the bulk dispersions along different routes marked in Fig. 3d, revealing the dispersions that have (Fig. 3e) and have not (Fig. 3f) degenerate points of ERs. These measured and simulated results consistently demonstrate the existence of ERs and drumhead bulk states in the non-Hermitian PC.

**Observation of DSSs at the surface of PC sample**

We then experimentally demonstrate the existence of DSSs at the surface perpendicular to the $z$ direction. In calculation, the PC ribbon has open boundaries in the $z$ direction and periodic boundaries in the $x$ and $y$ directions, which has 13 layers same as the cuboid-shaped PC sample. Figure 3g shows the projected dispersion along the $k_x$ direction with $k_y = 0$. The gray lines are the projected bulk states. The red lines represent the simulated DSSs near 5600 Hz in the bulk gap, which are confirmed by the measured data (color map). Figure 3h plots the measured drumhead surface dispersion at 5600 Hz on the $k_x$-$k_y$ plane, exhibiting the DSSs terminated by the projection of ERs (magenta curves). The experimental and simulated results have a good agreement with each other, demonstrating the presence of the DSSs.

**Observation of 3D GDSE in the finite-size PC samples**

We observe the 3D GDSE in the finite-size PC sample under pyramid-shaped geometry. Figure 4a shows the photograph of pyramid-shaped PC contained 9 layers, where the bottom and top layers possess $23 \times 23$ and $7 \times 7$ cavities, respectively. Four side surfaces are the same with each other because of the symmetry of lattice. We find that nonzero spectral winding numbers exist in the vertical direction of side surfaces, exhibiting spectral loop (nonzero spectral area) on the complex plane shown in Fig. 4b, which gives rise to skin effect on the side surfaces. Real part of spectra is plotted in Fig. 4c, where the color map represents the degree of localization at side surfaces and hinges, defined as $D = \sum_{x \epsilon L}|\psi(x)|^2$, and $L$ is the cavities in the outermost two layers. Strong localization is found at three frequency ranges, where the middle one is the drumhead surface states and the rest two are skin modes. We measure the pressure field distributions at 4750 Hz, 5250 Hz and 6200 Hz as shown in Figs. 4d-4f, respectively. Skin modes are observed at 5250 Hz and 6200 Hz, well consistent with the simulations in Fig. 4c. The skin effect emerges from the finite-size PC sample under pyramid-shaped geometry but disappears in that under cuboid-shaped geometry (Supplementary S-IV with Fig. S10 and S11), thus is the GDSE in three dimensions. The GDSE can also be revealed in the finite-size PC samples under other-shaped geometries (Supplementary S-V and S-VI with Fig. S12 and S13).

The GDSE in pyramid-shaped PC is a hybrid-order skin effect, which contains the surface and hinge skin modes simultaneously. To explore the surface skin modes, we build a PC ribbon whose open boundaries are the side surfaces of pyramid-shaped PC. Spectral area (brown dots) of PC ribbon is shown in Fig. 5a, which is different from that under periodic boundaries (gray dots), guaranteeing skin effect on the side surfaces. Top panel of Fig. 5b plots the spectra of pyramid-shaped PC, where the color map represents the degree of localization at side surfaces. One can see that some eigenstates at three frequency ranges perform as surface skin modes, which is demonstrated by the measured pressure response at surface (bottom panel), showing three resonance peaks larger than that at bulk. Surface skin mode is further evidenced by the measured pressure field distributions at 5125 Hz, as shown in Fig. 5c. It is noted that surface (bulk) response is measured at surface (bulk) cavity in $6^{th}$ layer, where a source is placed at surface (bulk) cavity in $4^{th}$ layer, as shown in Fig. 5c.

We finally investigate the hinge skin modes, which originates from the secondary deformation of the spectral area, similar to the higher-order skin effect in nonreciprocal systems [24]. Spectral area (green dots) of pyramid-shaped PC is shown in Fig. 5d, which is different from that of PC ribbon, giving rise to the secondary deformation.

This means that the surface skin modes will further localize and form the hinge skin modes. Top panel of Fig. 5e plots the spectra of pyramid-shaped PC, where the color map represents the degree of localization at hinges. Compared with surface skin modes, the hinge skin modes have a less number but possess stronger localization. This advantage makes the hinge skin mode easy to observe in the experiment. As shown in Fig. 5f, the measured pressure field is mainly confined to hinges, reviling the existence of hinge skin mode at 6000 Hz. We also measure the broadband response of hinge and plot it in the bottom panel of Fig. 5e, where the source and detector are placed at hinge (Fig. 5f). All the measured results are consistent with the spectra, demonstrating that the GDSE contains hinge skin modes.

**Discussion**

In conclusion, we have realized an exceptional line semimetal in a 3D non-Hermitian PC, which exhibits two types of bulk-boundary correspondence. Both the ERs and drumhead bulk states as the bulk features, and DSSs and 3D hybrid-order GDSE as the boundary properties, are observed ambiguously. Our work presents a complete non-Hermitian bulk-boundary correspondence of a 3D non-Hermitian topological semimetal, and may pave a way to potential applications in non-Hermitian topological devices for acoustic waves. Moreover, our system provides a guidance to further explore the non-Hermitian topological semimetals possessing non-Hermitian degeneracies with more complicated configurations, such as exceptional links, parabola and chains [2].

**References**
[1] Ashida, Y., Gong, Z. & Ueda, M. Non-Hermitian physics. Adv. Phys. **69**, 249-435 (2020).
[2] Bergholtz, E. J., Budich, J. C. & Kunst, F. K. Exceptional topology of non-Hermitian systems. Rev. Mod. Phys. **93**, 015005 (2021).
[3] Miri, M.-A. & Alù, A. Exceptional points in optics and photonics. Science **363**, eaar7709 (2019).
[4] Ding, K., Fang, C. & Ma, G. Non-Hermitian topology and exceptional-point geometries. Nat. Rev. Phys. **4**, 745-760 (2022).

[5] Kawabata, K., Bessho, T. & Sato, M. Classification of exceptional points and non-Hermitian topological semimetals. Phys. Rev. Lett. **123**, 066405 (2019).

[6] Kawabata, K., Shiozaki, K., Ueda, M. & Sato, M. Symmetry and topology in non-Hermitian physics. Phys. Rev. X **9**, 041015 (2019).

[7] Shen, H., Zhen, B. & Fu, L. Topological band theory for non-Hermitian Hamiltonians. Phys. Rev. Lett. **120**, 146402 (2018).

[8] Yang, Z., & Hu, J. Non-Hermitian Hopf-link exceptional line semimetals. Phys. Rev. B **99**, 081102(R) (2019).

[9] Carlström, J. & Bergholtz, E. J. Exceptional links and twisted Fermi ribbons in non-Hermitian systems. Phys. Rev. A **98**, 042114 (2018).

[10] Xu, Y., Wang, S.-T., & Duan, L.-M. Weyl exceptional rings in a three-dimensional dissipative cold atomic gas. Phys. Rev. Lett. **118**, 045701 (2017).

[11] Cerjan, A. et al. Experimental realization of a Weyl exceptional ring. Nat. Photonics **13**, 623-628 (2019).

[12] Liu, T., He, J. J., Yang, Z. & Nori, F. Higher-order Weyl-exceptional-ring semimetals. Phys. Rev. Lett. **127**, 196801 (2021).

[13] Liu, J. et al. Experimental realization of Weyl exceptional rings in a synthetic three-dimensional non-Hermitian phononic crystal. Phys. Rev. Lett. **129**, 084301 (2022).

[14] Wang, H., Ruan, J. & Zhang, H. Non-Hermitian nodal-line semimetals with an anomalous bulk-boundary correspondence. Phys. Rev. B **99**, 075130 (2019).

[15] Yang, Z., Zhang, K., Fang, C. & Hu, J. Non-Hermitian bulk-boundary correspondence and auxiliary generalized Brillouin zone theory. Phys. Rev. Lett. **125**, 226402 (2020).

[16] Yao, S. & Wang, Z. Edge states and topological invariants of non-Hermitian systems. Phys. Rev. Lett. **121**, 086803 (2018).

[17] Yokomizo, K. & Murakami, S. Non-Bloch band theory of non-Hermitian systems. Phys. Rev. Lett. **123**, 066404 (2019).

[18] Zhang, K., Yang, Z. & Fang, C. Correspondence between winding numbers and skin modes in non-Hermitian systems. Phys. Rev. Lett. **125**, 126402 (2020).

[19] Gong, Z. et al. Topological phases of non-Hermitian systems. Phys. Rev. X **8**, 031079 (2018).

[20] Okuma, N., Kawabata, K., Shiozaki, K. & Sato, M. Topological origin of non-Hermitian skin effects. Phys. Rev. Lett. **124**, 086801 (2020).

[21] Borgnia, D. S., Kruchkov, A. J. & Slager, R.-J. Non-Hermitian boundary modes and topology. Phys. Rev. Lett. **124**, 056802 (2020).

[22] Weidemann, S. et al. Topological funneling of light. Science **368**, 311-314 (2020).

[23] Zhang, L. et al. Acoustic non-Hermitian skin effect from twisted winding topology. Nat. Commun. **12**, 6297 (2021).

[24] Lee, C. H., Li, L. & Gong, J. Hybrid higher-order skin-topological modes in nonreciprocal systems. Phys. Rev. Lett. **123**, 016805 (2019).

[25] Kawabata, K., Sato, M. & Shiozaki, K. Higher-order non-Hermitian skin effect. Phys. Rev. B **102**, 205118 (2020).

[26] Zhu, W. & Gong, J. Photonic corner skin modes in non-Hermitian photonic crystals. Phys. Rev. B **108**, 035406 (2023).

[27] Zhang, X., Tian, Y., Jiang, J.-H., Lu, M.-H. & Chen, Y.-F. Observation of higher-order non-Hermitian skin effect. Nat. Commun. **12**, 5377 (2021).

[28] Zou, D. et al. Observation of hybrid higher-order skin-topological effect in non-Hermitian topolectrical circuits. Nat. Commun. **12**, 7201 (2021).

[29] Wang, W., Hu, M., Wang, X., Ma, G. & Ding, K. Experimental realization of geometry-dependent skin effect in a reciprocal two-dimensional Lattice. Phys. Rev. Lett. **131**, 207201 (2023).

[30] Wan, T., Zhang, K., Li, J., Yang, Z. & Yang, Z. Observation of the geometry-dependent skin effect and dynamical degeneracy splitting. Sci. Bull. **68**, 2330-2335 (2023).

[31] Zhou, Q. et al. Observation of geometry-dependent skin effect in non-Hermitian phononic crystals with exceptional points. Nat. Commun. **14**, 4569 (2023).

[32] Zhang, K., Yang, Z. & Fang, C. Universal non-Hermitian skin effect in two and higher dimensions. Nat. Commun. **13**, 2496 (2022).

[33] Zhang, X., Zhang, T., Lu, M.-H. & Chen, Y.-F. A review on non-Hermitian skin effect. Adv. Phys. X **7**, 2109431 (2022).

[34] Lin, R., Tai, T., Li, L. & Lee, C. H. Topological non-Hermitian skin effect. Front. Phys. **18**, 53605 (2023).

[35] Hu, B. et al. Non-Hermitian topological whispering gallery. Nature **597**, 655-659 (2021).

[36] Xue, H., Yang, Y. & Zhang, B. Topological acoustics. Nat. Rev. Mater. **7**, 974-990 (2022).

[37] Deng, W. et al. Nodal rings and drumhead surface states in phononic crystals. Nat. Commun. **10**, 1769 (2019).

[38] Zhou, H. et al. Observation of bulk Fermi arc and polarization half charge from paired exceptional points. Science **359**, 1009-1012 (2018).

## Methods

### Theoretical analysis

In Fig. 1d, we use spectral function to show the drumhead bulk states at $k_y = 0$, which is calculated by the formula $A(E) = -\frac{i}{\pi} \text{Im} G^r(E)$, where $G^r(E)$ and $E$ are the retarded Green function and the real part of eigenvalues of the tight-binding model, respectively. In Fig. 1e, we calculate the wave-function winding number with the intervals of $\Delta k_x = \Delta k_y = 2\pi/1000$. Under such point intervals, it can be numerically confirmed that the projections of ERs are the topological boundaries. In Figs. 2c and 2e, we calculate the spectral area of all eigenvalues of tight-binding Hamiltonian of the unit cell, where the point intervals are $\Delta k_x = \Delta k_y = \Delta k_z = 2\pi/100$.

### Numerical simulation

All the simulations are performed by the commercial COMSOL Multiphysics solver package, where the sound speed and air density are $346 \text{ m/s}$ and $1.29 \text{ kg/m}^3$, respectively. We choose the dipole mode along the $z$ direction of the acoustic cavity, so that the couplings along the $z$ direction are positive, and the couplings along the $x$ and $y$ directions are negative. The field distributions of eigenstates are obtained by extracting the sound pressure at the $h_0/4$ or $3h_0/4$ position of the acoustic cavity. In all simulation processes, the designed loss is induced by the imaginary part of sound speed with value $75i \text{ m/s}$.

### Experimental measurement

All the experimental PC samples in this work are fabricated by 3D printing technology. The wall thicknesses of these acoustic cavities and waveguides along the $x$ and $y$ directions are $2 \text{ mm}$, and the wall thicknesses of these waveguides along $z$ direction are $4 \text{ mm}$ because the sample was layered along this direction. At the same time, there are two rectangular holes with size $4 \times 6 \text{ mm}^2$ on the two side walls of those bigger waveguides along the $z$ direction, and we achieve the designed loss by stuffing sponges into these holes. To measure the sound pressure in all acoustic cavities, there are circular holes with a radius of 1.5 mm at $h_0/4$ and $3h_0/4$ positions of the surface acoustic cavities perpendicular to the $y$ direction of all samples. During the experiment, the sound source and detector are connected to the network analyzer (E5061B 5 Hz - 500 MHz) through wires. Response signals (forward transmission coefficient $S_{21}$) are measured at the concerned frequency range, and the frequency interval is 1.625 Hz. In Figs. 3c-3f, the point source was placed in the middle acoustic cavity of the cuboid-shaped PC for excitation, and by inserting detector into each acoustic cavities through these circular holes on the side cavities to measure the sound pressure of all these acoustic cavities, then the isofrequency curves and bulk dispersions are obtained by 3D Fourier transform. Similar to obtaining the nodal ring dispersions [37], the band dispersions are obtained by constructing the excited Bloch states. In Figs. 3g and 3h, we placed the sound source in the central acoustic cavity of the top surface of the cuboid-shaped PC to excite the DSSs, and by inserting detector into each acoustic cavities of upper surface to measure the sound pressure, then obtained the dispersion and isofrequency plane of DSSs by 2D Fourier transformation. In Figs. 4c, 4d and 5d-5f, we measure the pressure response by inserting the sound source and detector simultaneously into the $3h_0/4$ and $h_0/4$ positions for each acoustic cavity.

**ERs and drumhead bulk states**

Here, we show the ERs with opposite spectral winding numbers are connected by the drumhead bulk states. According to the Hamiltonian of the tight-binding model in Eq. (1), the eigenvalues can be obtained as

$$E = \pm\sqrt{(d_x^2 + d_y^2 - \gamma^2) - 2i\gamma(d_x \cos k_z + d_y \sin k_z)}. \tag{2}$$

Using Euler's formula, Eq. (2) has the form

$$E = \pm\sqrt{Re^{i\theta}} = \pm\sqrt{R}(\cos\frac{\theta}{2} + i\sin\frac{\theta}{2}), \tag{3}$$

where $R = \sqrt{Q_1^2 + Q_2^2}$, $Q_1 = R\cos\theta = d_x^2 + d_y^2 - \gamma^2$ and $Q_2 = R\sin\theta = -2\gamma(d_x \cos k_z + d_y \sin k_z)$.

The degeneracy of the real part of the eigenvalues requires $\cos\frac{\theta}{2} = 0$, i.e., $\theta = \pm\pi$, which means $Q_1 = -R \leq 0$ and $Q_2 = 0$, i.e.

$$d_x^2 + d_y^2 - \gamma^2 \leq 0, \tag{4}$$

$$2\gamma(d_x \cos k_z + d_y \sin k_z) = 0. \tag{5}$$

By substituting $d_x$ and $d_y$ into Eq. (5), we get

$$\cos k_x + \cos k_y = \frac{t_1 - t_2}{2t_0}\left(\frac{1}{\cos k_z} - \frac{2t_1}{t_1 - t_2}\cos k_z\right). \tag{6}$$

The Eq. (6) is further inserted into (4), obtaining

$$-\arctan\frac{\gamma}{t_2 - t_1} \leq k_z \leq \arctan\frac{\gamma}{t_2 - t_1}. \tag{7}$$

Both the Eqs. (6) and (7) describe the degeneracy of the real part of the eigenvalues. The existence of ERs requires the degeneracies of both the real and imaginary parts of eigenvalues, i.e., $Q_1 = Q_2 = 0$, obtaining $k_z = \pm \arctan\frac{\gamma}{t_2 - t_1}$. So there are a pair of ERs located at $k_z = \arctan\frac{\gamma}{t_2 - t_1}$ and $k_z = -\arctan\frac{\gamma}{t_2 - t_1}$, which possess spectral winding number $\pm 1$ calculated in Supplementary S-I C. It can be seen from Eq. (7) that ERs with opposite spectral winding numbers are connected by the drumhead bulk states that have only the degeneracy of the real part of the eigenvalues.

When $\gamma = 0$ for Hermitian case, the eigenvalues belong to real domain. From Eq. (7), the degeneracy of the eigenvalues requires $k_z = 0$ (or $\pi$), forming a nodal ring on the $k_z = 0$ plane, as shown by the black dashed curve in Fig. 1b.

**Data availability**

The data that support the plots within this paper and other findings of this study are available from the corresponding author upon reasonable request.

**Acknowledgements**

We thank Zhesen Yang for helpful discussions. This work is supported by the National Key R&D Program of China (Nos. 2022YFA1404500, 2022YFA1404900), National Natural Science Foundation of China (Nos. 12074128, 12222405, 12374409), and Guangdong Basic and Applied Basic Research Foundation (Nos. 2021B1515020086, 2022B1515020102).

**Author contributions**

All authors contributed extensively to the work presented in this paper.

**Competing financial interests**

The authors declare no competing financial interests.

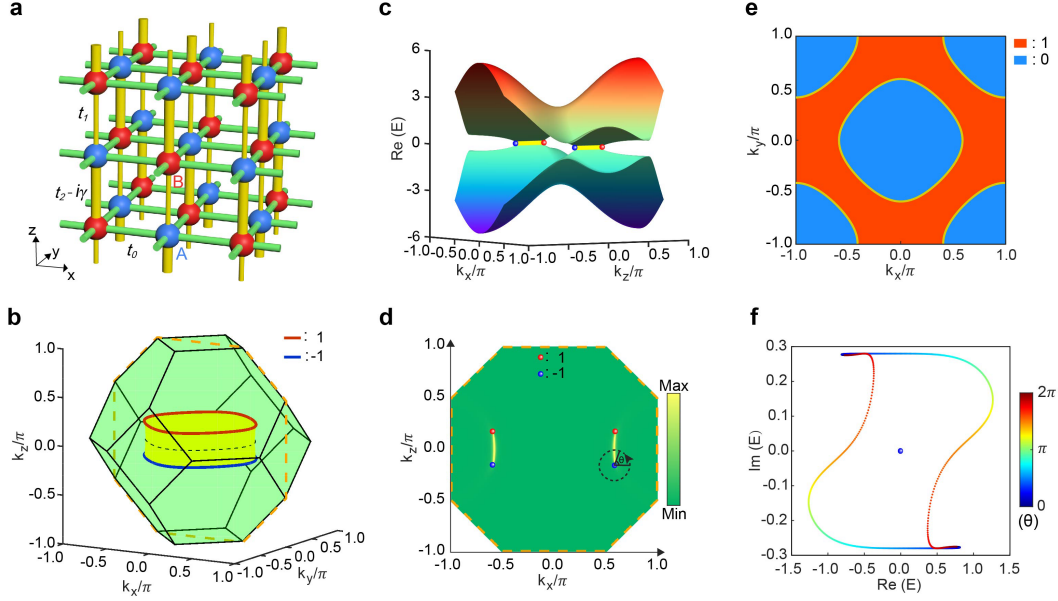

**Fig. 1 | Exceptional line semimetal hosting nontrivial non-Hermitian bulk topologies in a tight-binding model. a**, Schematics of the lattice model. Unit cell contains two sites A and B. The intralayer couplings are $t_0$ and the interlayer couplings are $t_1$ and $t_2 - i\gamma$. Non-Hermiticity is induced by the loss with strength $\gamma$. **b**, Distribution of paired ERs with $\pm 1$ spectral winding numbers (red and blue solid curves) connected by drumhead bulk states (yellow area) in the first BZ (green truncated octahedron). The pair of ERs are evolved from a nodal ring (black dashed curve) in the Hermitian case with $\gamma = 0$. **c,d**, Bulk dispersion and isofrequency curve of zero energy at $k_y = 0$ plane of the first BZ enclosed by orange dashed lines in **b**, showing a cross section of ERs (red and blue dots) and drumhead bulk states (yellow lines). **e**, Distribution of wave-function winding number $w$ originated from wave-function topology of ERs on the $k_x$-$k_y$ plane. **f**, Spectral loop on the complex plane along the path of the black dashed circle in **d**, revealing spectral winding number $v = -1$ attributed to spectral topology of ER. The parameters are chosen as $t_0 = -1$, $t_1 = 0.5$, $t_2 = 1$, $\gamma = 0.28$ in calculation.

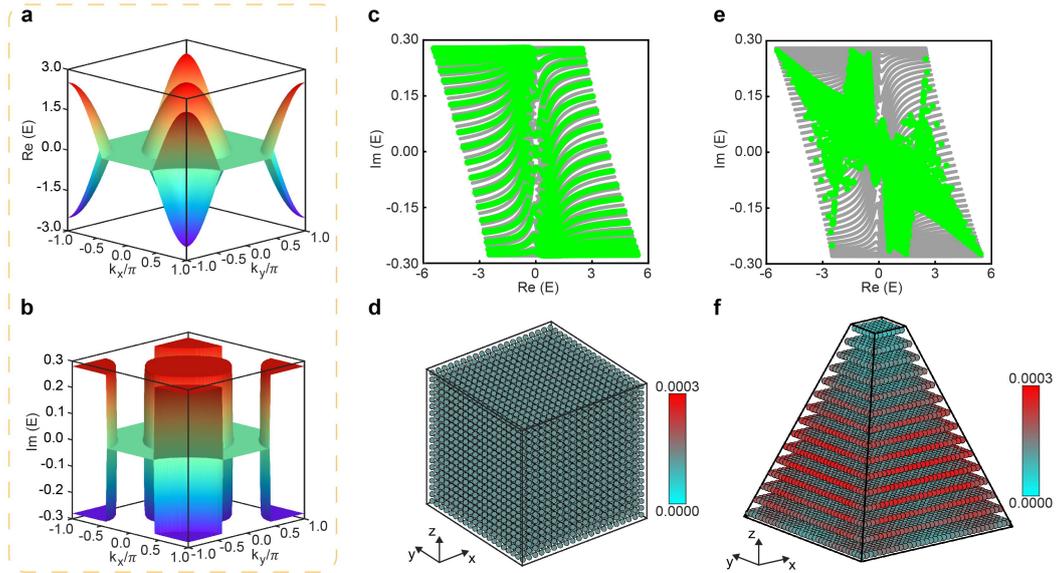

**Fig. 2 | Non-Hermitian bulk-boundary correspondence in the exceptional line semimetal. a,b**, Real and imaginary parts of drumhead surface dispersion calculated by a ribbon sample. The DSSs are localized at the open boundaries in the $z$ direction. **c**, Spectral area (green dots) under cuboid-shaped geometry with fully open boundaries. Gray dots represent the spectral area under periodic boundaries. **d**, Spatial distribution of eigenstates under cuboid-shaped geometry. **e,f**, The same to **c,d** but under pyramid-shaped geometry. Skin effect occurs in **f** but not in **d**, revealing the GDSE in the 3D sample.

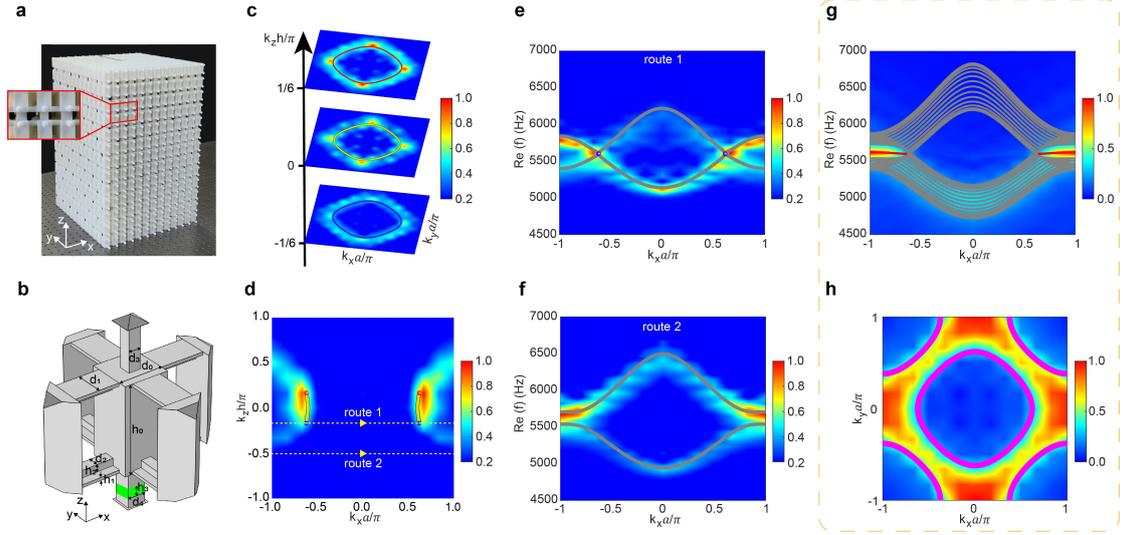

**Fig. 3 | ERs and drumhead bulk states, and DSSs in the non-Hermitian PC. a**, Photograph of the 3D non-Hermitian sample with an enlarge front view shown in the inset. The designed loss is achieved by the holes on bigger interlayer waveguide sealed with the sound-absorbing sponges (black). **b**, Schematic of the PC unit cell. Green areas represent the rectangular holes on the bigger waveguide along the $z$ direction. **c**, Measured (color map) and simulated (colored curves) isofrequency curves at 5600 Hz on the $k_z = \pm\pi/(6h)$, 0 planes, showing ERs and partial drumhead bulk states, respectively. **d**, Measured and simulated isofrequency curves on the $k_y = 0$ plane, presenting the drumhead bulk states terminated by ERs. **e**,**f**, Measured (color map) and simulated (gray lines) bulk dispersions for routes 1 and 2 marked in **d**. **g**, Projected dispersions of DSSs. Color map denotes the measured results, while red and gray lines represent the simulated DSSs and bulk states, respectively. **h**, Measured drumhead surface dispersion at 5600 Hz. Magenta curves show the projected ERs on the $k_x$-$k_y$ plane.

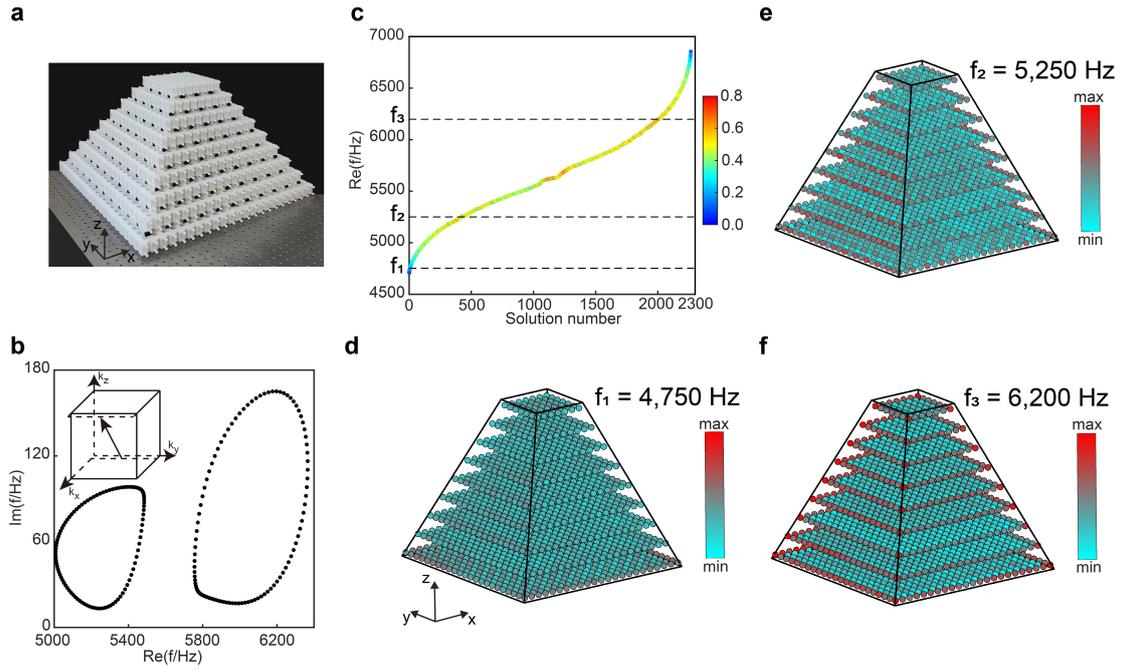

**Fig. 4 | Observation of GDSE in the finite-size PC sample under pyramid-shaped geometry. a**, Photograph of the pyramid-shaped PC sample with 9 layers. **b**, Spectral loop on the complex plane along the vertical direction of side surface (inset), revealing nonzero spectral winding number. **c**, Real part of spectra for pyramid-shaped PC sample. The color represents the degree of localization for each eigenstate on side surfaces and hinges. **d-f**, Measured pressure field distributions at $f_1 = 4750$ Hz, $f_2 = 5250$ Hz and $f_3 = 6200$ Hz, respectively. The skin effect is obviously observed at $f_2$ and $f_3$.

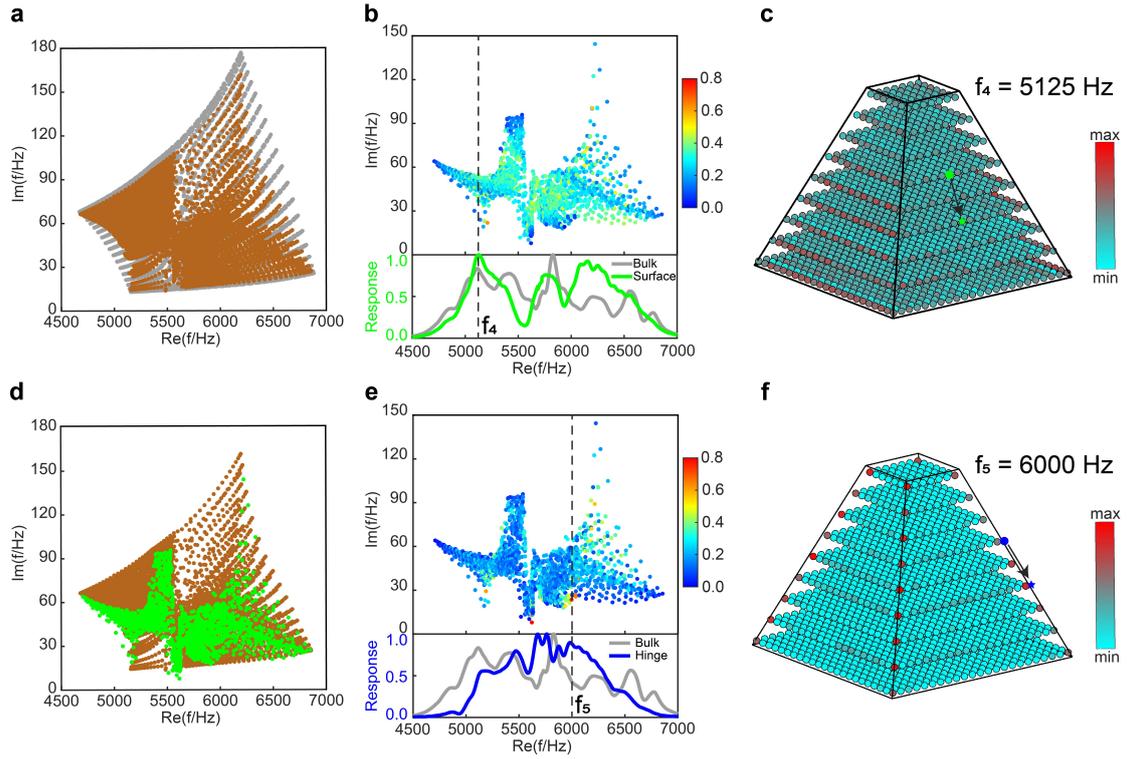

**Fig. 5 | Observation of hybrid-order feature of GDSE. a**, Spectral area (brown dots) of PC ribbon whose open boundaries are the side surfaces of pyramid-shaped PC. Gray dots represent the spectral area of PC under periodic boundaries. **b**, Top panel: spectral area of pyramid-shaped PC, where the color represents the degree of localization for each eigenstate on four side surfaces. Bottom panel: measured response spectra at bulk and surface. **c**, Measured pressure field distributions at $f_4 = 5125$ Hz, showing surface skin modes. **d**, Spectral area (green dots) of pyramid-shaped PC. **e**, Top panel: spectral area of pyramid-shaped PC, where the color represents the degree of localization for each eigenstate on four side hinges. Bottom panel: measured response spectra at bulk and hinge. **f**, Measured pressure field distributions at $f_5 = 6000$ Hz, revealing hinge skin modes.